\documentstyle[12pt,aaspp4,epsf]{article}
\def\ltsim{ \,{}^<_\sim\, }

\begin{document}

\title{The Hubble Constant From observations of
\\ the Brightest Red Giant Stars in a Virgo-cluster Galaxy}

\author{William E.~Harris}
\affil{Department of Physics \& Astronomy, McMaster University 
\\Hamilton, ON L8S 4M1, Canada
\\harris@physics.mcmaster.ca}

\author{Patrick R.~Durrell}
\affil{Department of Astronomy, Case Western Reserve University
\\Cleveland, OH 44106, USA 
\\durrell@huascaran.astr.cwru.edu}

\author{Michael J.~Pierce}
\affil{Department of Astronomy, Indiana University 
\\Bloomington, IN 47405, USA
\\mpierce@astro.indiana.edu}

\author{Jeff Secker}
\affil{Program in Astronomy, Washington State University 
\\Pullman, WA 99164-3113, USA
\\secker@delta.math.wsu.edu}

\bigskip
\centerline{June 4, 1998}

\bigskip \bigskip
{\bf 
The nearest large groups of elliptical galaxies, in the Virgo 
and Fornax clusters,
play central roles in determining the Hubble constant
$H_0$ and thus the cosmological rate of expansion.
Since the relative distances between these two clusters and more
remote clusters are well known, robust distance determinations
to Virgo and Fornax will establish the Hubble constant for the
local universe.  In addition, elliptical galaxies reside 
predominantly in the cores of clusters, so that distance
calibrations for ellipticals will minimize the uncertainties
due to the possible large extent of clusters along the line of sight.
A powerful and direct way of establishing such distances
is to use the brightest red-giant stars, which have nearly
uniform luminosities$^{1,2}$.  Here we 
report the direct observation of the old red giant 
stars in a dwarf elliptical galaxy in the Virgo cluster.  
We determine a distance to this galaxy, and thus to the core
of the Virgo cluster, of $\sim 15.7 \pm 1.5$
Megaparsecs, from which we estimate a Hubble constant of 
$H_0 = 77 \pm 8$ km s$^{-1}$ Mpc$^{-1}$.  
Under the assumption of a low-density Universe with the simplest
cosmology, the age of the Universe is no more than 12-13 billion years.
}

\bigskip

For the old, low-mass stars that dominate the stellar population in
elliptical galaxies, the post-main-sequence evolution along
the giant branch terminates at the point of core helium ignition
which defines the tip of the red giant branch (TRGB). If these 
stars are imaged in the near-infrared $I$ band, the small dependence
of the TRGB bolometric luminosity on stellar composition (metallicity)
almost completely cancels the bolometric
correction, leaving the absolute magnitude $M_I$(TRGB) at the same
level to within $\pm0.1$ magnitude for any metallicity in the
range [Fe/H] $\ltsim -0.7^{1,2,3,4}$.  The TRGB luminosity can
be calibrated from the giant branches in nearby    
globular clusters$^{1,5}$, whose distances are set
in turn by RR Lyrae parallaxes and main-sequence fitting to subdwarfs.
The TRGB method thus yields distance measurements for other
galaxies that are only two steps removed from fundamental
trigonometric parallaxes and independent of any other
methods involving Local Group calibrators.
By contrast, the Cepheid-based distance calibrations$^{6,7,8,9,10}$
lie three or four steps away from fundamental parallax methods by the time
they reach remote galaxies such as the rich proving grounds of
Virgo and Fornax. 

We have used the superb imaging resolution of the 
{\it Hubble Space Telescope} to achieve the first TRGB
distance calibration to a Virgo galaxy.
Our target was the nucleated dwarf elliptical
VCC 1104 = IC 3388, which has a projected location $43'$ 
(or $\sim 200$ kpc) northeast of the central giant M87,
and thus well within the $\sim 1$-Megaparsec projected core radius
of the Virgo ellipticals$^{11}$.
We obtained deep $I-$band (filter F814W) exposures
of this galaxy in June 1997 with the WFPC2 camera on HST.  
Twelve images adding up to a total exposure time of 32200 sec were 
re-registered and co-added
to generate a final composite image in which the galaxy is 
beautifully resolved, displaying a sheet of faint stars across the entire 
WFPC2 field.

Photometry of the stars on the final image was carried out 
with the DAOPHOT II suite of software$^{12}$.  
To set the zero point of the $I-$magnitude scale, we used
our ground-based WIYN photometry to determine $I$
magnitudes for 6 of the brightest stars in the WFPC2 image field,
and then determined the mean offset between the instrumental
magnitudes and the true $I$ to within 
an uncertainty $\sigma(I) = \pm0.04$.  The normal prescription
for self-calibration of the WFPC2 F814W filter via aperture 
photometry$^{13}$ was also used on 14 stars in the WF2,3,4 fields, yielding
a zeropoint with $\sigma(I) = \pm 0.03$ uncertainty.  These two methods
gave zeropoints which differed by only 0.02 mag.   An additional systematic
zeropoint uncertainty may arise when we transfer the magnitude scale for
the relatively bright calibration stars to the faint RGB stars we are 
interested in, because of the charge transfer efficiency effect on the WFPC2
CCDs$^{14}$.  The CTE effect should, however,
be relatively small here since our individual
(full-orbit) exposures are the longest possible, yielding background
levels of 25 DN per exposure.  The published CTE equations$^{14}$ suggest 
that any zeropoint correction is likely to be no larger than $\simeq 0.04$ mag,
but it should be kept in mind when interpreting the total error budget
listed in Table 1.

We used additional ground-based images of VCC 1104 (obtained with
the WIYN telescope at Kitt Peak National 
Observatory) to measure its integrated 
color profiles and thus its metallicity.  We find $(U-B) = 0.20 
\pm 0.05$ and $(B-V) = 0.74 \pm 0.02$, which give [Fe/H] $\simeq -1.3
\pm 0.2$ through the normal relation between metallicity and integrated 
color for globular clusters$^{15}$.  The galaxy is therefore squarely in 
the low-metallicity regime within which $M_I$(TRGB) is constant.

Close to the center of VCC 1104, individual stars are too crowded 
for reliable photometry even with HST's high resolution. 
We conservatively kept only measurements of stars more than $15''$
(equivalent to about 1 kiloparsec in projected distance) 
from the galaxy center, leaving a total `clean' sample of $\simeq 1500$ 
stars brighter than $I = 27.5$ for which blending and crowding 
were negligible.  Extensive artificial-star simulations show that 
our limiting magnitude (50\% detection completeness) is 
$I \simeq 27.5$ in these outer regions.  

The raw luminosity function (LF) of the detected stars 
is shown in the upper panel of Fig.~1.
It has three components:  (a) the dominant contribution
of the RGB stars in the galaxy,
(b) a small ($\ltsim 20$\%) 
contribution of brighter asymptotic giant stars (AGB)
on their second GB ascent, and (c) a 
contribution from faint, starlike background objects.
We model the entire LF 
through a maximum-likelihood fitting approach$^{16}$.
The RGB luminosity function $N(I)$ is assumed to have a step-function
onset at $I$(TRGB) followed by an exponential rise to fainter magnitudes.
To this we add an AGB population starting 1.0 mag brighter than the 
TRGB and rising exponentially to fainter magnitudes (though with 
much smaller amplitude than the RGB itself).  We set the exponential slopes
for both RGB and AGB at $\Delta$log N/$\Delta I$ = 0.96
from the (log $N$, $I$) form of the LF.
Finally, we convolve the ideal model with  
the photometric error function $\sigma(I)$ and 
completeness function $f(I)$ (both of which are known from the artificial-star 
simulations). The smoothed background $b(I)$ is added,
and the fully transformed model is finally matched to the observed LF.  
In general, the LF for this galaxy very strongly resembles the ones
published recently for two dwarf ellipticals in the nearby M81 
group$^{17}$, for which the $(I,V-I)$ color-magnitude diagrams
exhibit a strong RGB component terminating sharply at the
TRGB, with a sprinkling of AGB stars continuing upward brighter
than the tip for about 0.8 magnitude.

The free parameters in the LF fit are the 
assumed $I$(TRGB) and the fractional contribution of the AGB, which are
adjusted to find the best match (maximum likelihood) to the normalized LF.  
Our model solution gives $I$(TRGB) = $26.82 \pm 0.06$.
Variations in the size of the AGB population, or differences in the
RGB slope exponent, do affect the overall {\it goodness of fit}
of the model LF, but the deduced {\it magnitude of
the TRGB} is almost completely insensitive to either parameter because
of its intrinsically sharp step-function nature.  (Realistic
changes in the exponential slope, or the AGB population ratio, resulted
in only $\pm0.01-0.02$ differences in the TRGB value). 
Similarly, even neglecting the contribution of the background $b(I)$ 
altogether does not change the deduced TRGB significantly.

As a check on the solution, we applied the Sobel edge-detection 
filter$^{1,2}$ in its continuous form to the observed luminosity 
function (lower panel of Fig.~1).  This method, though cruder,
has the advantage that it is model-independent; it gives the
same answer for $I$(TRGB) to within $\pm0.1$ magnitude.

The distance to the galaxy is set once we adopt an
absolute magnitude for the TRGB, $M_I$(TRGB) $ = -4.2 \pm
0.1$ (see Fig.~2).  Our derived distance modulus for VCC 1104 is then
$(m-M)_0 = 30.98 \pm 0.20$ for a foreground Virgo absorption $A_I = 0.04$.  
Our resulting distance to VCC 1104, and therefore the Virgo
cluster core, is 
$$ d(Virgo) = (15.7 \pm 1.5)~ {\rm Mpc}.  $$
This result agrees to within its quoted uncertainty
with other recent measurements such
as through planetary nebula luminosity functions$^{18}$,
surface brightness fluctuations for E/S0 galaxies$^{19}$,
and the Cepheids in spiral galaxies that are high-probability
Virgo members$^{6,7,8,9,10}$.

The error budget is summarized in Table 1 along with those for
the Cepheid  and PNLF (planetary nebula) techniques.  
For single Virgo galaxies selected {\it at random},
a major advantage of either the TRGB or PNLF 
methods should be a smaller geometric
depth effect, since the Virgo `core' defined by the elliptical
galaxies is expected to be smaller than the more widespread Cepheid
spirals (but see below).  

Two extreme possibilities may have significantly biased 
our result: (1) We may have unluckily picked a target which genuinely 
lies on the near (or far) side of the Virgo cluster core.  
The distribution of intergalactic planetary
nebula within Virgo$^{20}$ suggests that the intracluster {\it stars}
may extend as much as 3 Mpc towards us along the line of sight,
though it is unclear whether or not the core distribution of
E and dE {\it galaxies} is similarly extended$^{11}$.
Ultimately, the only safeguard we have against this type of accident is to 
acquire data for more galaxies of similar type and determine their 
distribution of distances.  (2) This particular dwarf elliptical
could have a huge population of bright AGB stars which mimic 
the normal first-ascent RGB tip, and which would cause us to 
underestimate its distance. However, no such AGB population is seen 
in the outskirts of similar dwarf ellipticals (see ref.~17 for an
excellent comparison),
or even in the halos of larger galaxies$^{21,22,23}$.  

To compute the Hubble constant, we use our measured Virgo distance
to step outward to the Coma cluster, whose mean redshift in the
CMB frame$^{24,25}$ is $v(Coma) = (7100 \pm 100)$ km s$^{-1}$.
The relative distance modulus $\Delta m$(Coma--Virgo) appears
now to be the most poorly known quantity in the logical chain,
with measured estimates ranging from $\simeq 3.4$ to 4.1$^{26}$.
Recent measurements particularly through the $D_n-\sigma$ method
for fundamental-plane ellipticals$^{27,28}$ suggest a mean
$\langle \Delta m \rangle \simeq 3.85 \pm 0.15$, which we adopt here.
Thus, $d$(Coma) = $(92.5 \pm 10.3)$ Mpc and the Hubble constant is
$$ H_0 = (77 \pm 9)~ {\rm km}~ s^{-1}~ {\rm Mpc}^{-1} . $$
The corresponding expansion age for the universe is
$(12.6 \pm 1.5)$ Gyr for $\Omega \simeq 0$, or $(8.4 \pm 1.0)$ Gyr
for critical density ($\Omega = 1$).  The former value, for a very
low-density universe, now fits with little room to spare within 
the lowest currently estimated
ages for the oldest Milky Way globular clusters, which are 
in the range of $11 - 13$ Gyr$^{29,30}$.  The high$-\Omega$ expansion
time is, of course, still strongly in conflict with these ages.  

Correspondence and requests for materials should be directed to 
W. E. Harris (harris@physics.mcmaster.ca).

\clearpage

\noindent {\bf References}
\bigskip
\parindent = -5 true mm

1.  Lee, M.~G., Freedman, W.~L., \& Madore, B.~F.
The Tip of the Red Giant Branch As a Distance Indicator for
Resolved Galaxies.
{\it Astrophys.J.} {\bf 417}, 553-559 (1993).

2. Mould, J., \& Kristian, J.  The Stellar Populations in the
Halos of M31 and M33.  {\it Astrophys.J.} {\bf 305}, 591-599 (1986).

3.  Lee, M.~G.  The Distance to Nearby Galaxy NGC 3109 Based on
the Tip of the Red Giant Branch.  {\it Astrophys.J} {\bf 408},
409-415 (1993).

4.  Sakai, S., Madore, B.~F., \& Freedman, W.~L.
Tip of the Red Giant Branch Distances to Galaxies. III.
The Dwarf Galaxy Sextans A.
{\it Astrophys.J.} {\bf 461}, 713-723 (1996).

5.  Da Costa, G.~S., \& Armandroff, T.~E.  Standard Globular Cluster
Giant Branches in the ($M_I$, $(V-I)_0$) Plane.  
{\it Astron.J.} {\bf 100}, 162-181 (1990).  

6.  Pierce, M.~J., Welch, D.~L., McClure, R.~D., van den Bergh, S.,
Racine, R., \& Stetson, P.~B.  The Hubble Constant and Virgo Cluster
Distance from Observations of Cepheid Variables.  {\it Nature}
{\bf 371}, 385-389 (1994).

7. Freedman, W.~L. et al.  Distance to the Virgo Cluster galaxy
M100 from Hubble Space Telescope Observations of Cepheids.
{\it Nature} {\bf 371}, 757-762 (1994).

8.  Ferrarese, L. et al.  The Extragalactic Distance Scale Key 
Project.  IV.  The Discovery of Cepheids and a New Distance
to M100 Using the HST.  {\it Astrophys.J.} {\bf 464}, 568-599 (1996).

9.  Saha, A., Sandage, A., Labhardt, L., Tammann, G.~A.,
Macchetto, F.~D., \& Panagia, N.  Cepheid Calibration of the
Peak Brightness of SNeIa.  V.   SN1981B in NGC 4536.
{\it Astrophys.J.} {\bf 466}, 55-91 (1996).

10.  Saha, A., Sandage, A., Labhardt, L., Tammann, G.~A.,
Macchetto, F.~D., \& Panagia, N.  Cepheid Calibration of the
Peak Brightness of SNeIa.  VI.  SN1960F in NGC 4496A.
{\it Astrophys.J.Suppl.} {\bf 107}, 693-737 (1996).

11.  Ferguson, H.~C., \& Sandage, A. The Spatial Distributions and
Intrinsic Shapes of Dwarf Elliptical Galaxies in the Virgo and
Fornax Clusters.
{\it Astrophys.J.Letters} {\bf 346}, L53-L56 (1989).

12.  Stetson, P.~B., Davis, L.~E., \& Crabtree, D.~R.  Future Development
of the DAOPHOT Crowded-Field Photometry Package, in 
{\it CCDs in Astronomy}, ASP Conf.~Ser.~{\bf 8},
edited by G.~H.~Jacoby (ASP, San Francisco), 289-304 (1990)

13.  Holtzman, J.~A. et al.  The Photometric Performance and Calibration
of WFPC2.  {\it Publ.Astron.Soc.Pacific} {\bf 107}, 1065-1093 (1995).

14.  Whitmore, B., \& Heyer, I. New Results on Charge Transfer Efficiency
and Constraints on Flat-Field Accuracy.  Space Telescope Science Institute,
Instrument Science Report 97-08 (1997).

15. Couture, J., Harris, W.~E., \& Allwright, J.~W.~B.
BVI Photometry of Globular Clusters in M87. {\it Astrophys.J.Suppl.}
{\bf 73}, 671-683 (1990).

16.  Secker, J., \& Harris, W.~E.  A Maximum-Likelihood Analysis
of Globular Cluster Luminosity Distributions in the Virgo Cluster.
{\it Astron.J.} {\bf 105}, 1358-1368 (1993).

17.  Caldwell, N., Armandroff, T.~E., Da Costa, G.~S., \& Seitzer,
P.  Dwarf Elliptical Galaxies in the M81 Group:  The Structure
and Stellar Populations of BK5N and F8D1.  {\it Astron.J.}
{\bf 115}, 535-558 (1998).

18.  Jacoby, G.~H., Ciardullo, R., \& Ford, H.~C.
Planetary Nebulae as Standard Candles. V.  The Distance to the
Virgo Cluster.
{\it Astrophys.J.} {\bf 356}, 332-349 (1990).

19.  Tonry, J.~L., Blakeslee, J.~P., Ajhar, E.~A., \& Dressler, A.
The SBF Survey of Galaxy Distances. I.  Sample Selection, Photometric
Calibration, and the Hubble Constant. 
{\it Astrophys.J.} {\bf 475}, 399-413 (1997).

20.  Feldmeier, J.~J., Ciardullo, R., \& Jacoby, G.~H.
Intracluster Planetary Nebulae in the Virgo Cluster I.  Initial Results.
{\it Astrophys.J.} in press (1998)

21.  Soria, R. et al.  Detection of the Tip of the Red Giant
Branch in NGC 5128.
{\it Astrophys.J.} {\bf 465}, 79-90 (1996).

22.  Elson, R.~A.~W. Red giants in the halo of the S0 galaxy NGC 3115: a
distance and a bimodal metallicity distribution.
{\it Mon.Not.R.Astron.Soc.} {\bf 286}, 771-776 (1997).

23.  Sakai, S., Madore, B.~F., Freedman, W.~L., Lauer, T.~R.,
Ajhar, E.~A., \& Baum, W.~A. Detection of the Tip of the
Red Giant Branch in NGC 3379 (M105) in the Leo I Group Using
the Hubble Space Telescope.
{\it Astrophys.J.} {\bf 478}, 49-57 (1997).

24.  Colless, M., \& Dunn, A.~M.  Structure and Dynamics of the
Coma Cluster.
{\it Astrophys.J.} {\bf 458}, 435-454 (1996).

25.  Mould, J. et al.  Limits on the Hubble Constant from the
HST Distance of M100. 
{\it Astrophys.J.} {\bf 449}, 413-421 (1995).

26.  van den Bergh, S.  The Hubble Parameter.
{\it Publ.Astron.Soc.Pacific} {\bf 104}, 861-883 (1992).

27.  Faber, S.~M. et al.  Spectroscopy and Photometry of Elliptical
Galaxies:  VI.  Sample Selection and Data Summary.
{\it Astrophys.J.Suppl.} {\bf 69}, 763-808 (1989).

28.  Hjorth, J., \& Tanvir, N.~R.  Calibration of the Fundamental
Plane Zero Point in the Leo I Group and an Estimate of the
Hubble Constant.  
{\it Astrophys.J.} {\bf 482}, 68-74 (1997).

29.  Harris, W.~E. et al.  NGC 2419, M92, and the Age Gradient
in the Galactic Halo.  {\it Astron.J.} {\bf 114}, 1030-1042 (1997).

30.  Gratton, R.~G., Fusi Pecci, F., Carretta, E., Clementini, G.,
Corsi, C.~E., \& Lattanzi, M.  Ages of Globular Clusters from
Hipparcos Parallaxes of Local Subdwarfs.
{\it Astrophys.J.} {\bf 491}, 749-771 (1997).

31.  Cassisi, S., \& Salaris, M.  The tip of the Red Giant Branch as
a Distance Indicator:  theoretical calibration and the value of $H_0$.
{\it Mem.soc.Astron.Ital.}, in press (1998).

\clearpage

\begin{deluxetable}{lcc}
\vsize 9.5 true in
\tablenum{1}
\tablecaption{Virgo Distance Error Budget \label{tab1}}
\tablewidth{0pt}
\tablehead{
\colhead{Source} & \colhead{$\sigma(m-M)$ (magnitudes)}
}
\startdata
{\bf TRGB:} & \nl
WFPC2 photometric zeropoint uncertainty & $\pm 0.03$  \nl
TRGB uncertainty from LF model fit & $\pm 0.06$  \nl
Uncertainty in Virgo foreground extinction & $\pm 0.02$  \nl
Internal scatter in M$_{I}$(TRGB) calibration & $\pm 0.10$  \nl
Uncertainty in globular cluster distance scale & $\pm 0.10$  \nl
Geometric depth of Virgo core ($\pm 1$ Mpc) & $\pm 0.13$  \nl
 \nl
{\bf Total uncertainty for TRGB technique (1 galaxy)} & $\pm 0.20$  \nl
 \nl
{\bf Cepheids:} & \nl
{\it P-L} Relation fit of LMC to Virgo spiral & $\pm 0.17$ \nl
Uncertainty in Cepheid metallicity & $\pm 0.05$ \nl
Uncertainty in adopted LMC distance & $\pm 0.10$ \nl
Uncertainty in Virgo extinction relative to LMC & $\pm 0.02$ \nl
Geometric depth of Virgo spirals ($\pm 3$ Mpc) & $\pm0.35$ \nl
 \nl
{\bf Total uncertainty for Cepheid technique (1 galaxy)} & $\pm 0.41$ \nl
\nl
{\bf Planetary Nebulae:} & \nl
Fitting uncertainty to observed PNLF & $\pm 0.10$ \nl
Photometric zeropoint and filter calibration uncertainties & $\pm 0.06$
\nl
Uncertainty in foreground Virgo extinction relative to M31 & $\pm 0.02$
\nl
Uncertainty in fiducial M31 distance & $\pm 0.10$ \nl
Definition in fiducial model PNLF & $ \pm 0.05$ \nl
Geometric depth of Virgo core ($\pm 1$ Mpc) & $\pm 0.13$ \nl
 \nl
{\bf Total uncertainty for PNLF technique (1 galaxy)} & $\pm 0.21$ \nl

\enddata
\tablenotetext{}{A quantitative overview of the principal internal and
external uncertainties in determining the Virgo distance modulus.
Figures shown are for {\it one} Virgo galaxy picked {\it at random},
either
from the Virgo E and dE population (for the TRGB and PNLF techniques), or from
the spirals (for the Cepheid technique).  Data for the PNLF technique
are taken from ref.~18, and from ref.~8 for the Cepheids.
}
 
\end{deluxetable}
\clearpage

\figcaption[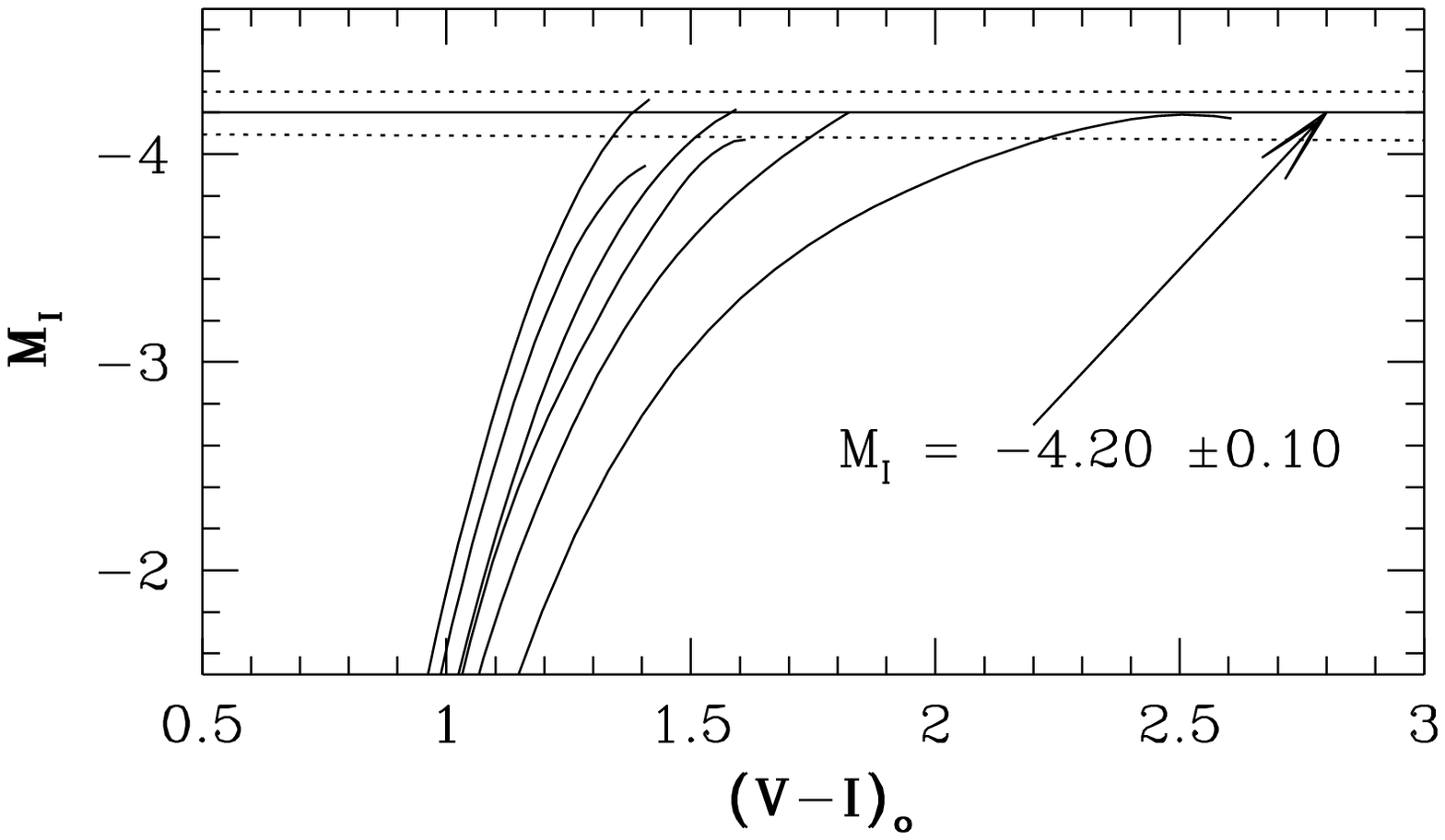]{
{\it  Upper panel:} The observed luminosity function 
$N(I)$ (number of stars per 0.1-magnitude bin) 
for the red giants in VCC 1104. The solid line 
superimposed on the raw data shows the scaled maximum-likelihood 
fit of the model LF to the data, scaled to match the total 
population of stars and including the effects of background 
contamination, photometric error, and detection completeness. The steep 
rise starting near $I \simeq 26.7$ shows the tip of the red 
giant branch; ideally, it would show up as a step-function jump 
exactly at the RGB tip, but it is broadened out by the photometric 
measurement uncertainty, which is $\sigma (I) = \pm 0.15$ mag at 
the TRGB level and increases steadily to fainter levels. The steep falloff 
past $I \sim 27.4$ is due to the declining completeness of detection 
of faint stars, which is nearly 100\% at the TRGB but 
drops to 50\% at $I \simeq 27.5$ and effectively to zero at 
$I = 28$. The small bump in the LF at the bright end 
($I \simeq 26.3$) is likely to be due to the presence of
bright AGB stars. The local background function 
$b$ -- the raw, unsmoothed form of which
is shown as the hatched histogram -- is defined 
empirically from the LF on 
the outermost half of the WF4 camera field, which is furthest
from the center of VCC 1104 and in which the stellar density is lowest.
However, even at the outer edges of the field, the RGB population
from the galaxy contributes noticeably to $b$ and thus the ``true''
background level is significantly lower than is shown here.
{\it  Lower panel:} Results of the Sobel 
edge-detection algorithm, as applied to the smoothed and 
normalized LF. The function $E(I)$ shows a sharp peak wherever 
it passes through a rapid change in the LF. The AGB population
shows up as a small ``edge'' at $I \sim 26.3$, but it is the 
peak at $I \sim 26.7$ which we identify as the RGB tip. }

\figcaption[Harris.fig3.ps]{
Fiducial giant branches for standard Milky Way globular 
clusters (absolute magnitude $M_I$ against color $(V-I)_0$). The 
clusters measured by Da Costa and Armandroff (ref.~5) have been 
recalibrated through the Hipparcos subdwarf parallax distance 
scale (ref.~30). The sequence of cluster RGB's shown covers the metallicity 
range [Fe/H] = $-2.2$ for M15 (leftmost line) up to [Fe/H] $= -0.7$ 
for 47 Tucanae (rightmost line); over this range, the 
red giant branch tip is virtually constant 
at $M_I$(TRGB) $= -4.2 \pm 0.1$.  Encouragingly,
recent theoretical RGB models (ref.~31) agree well with the empirical
calibration.}

\clearpage
\begin{figure}
\epsscale{1.0}
\plotone{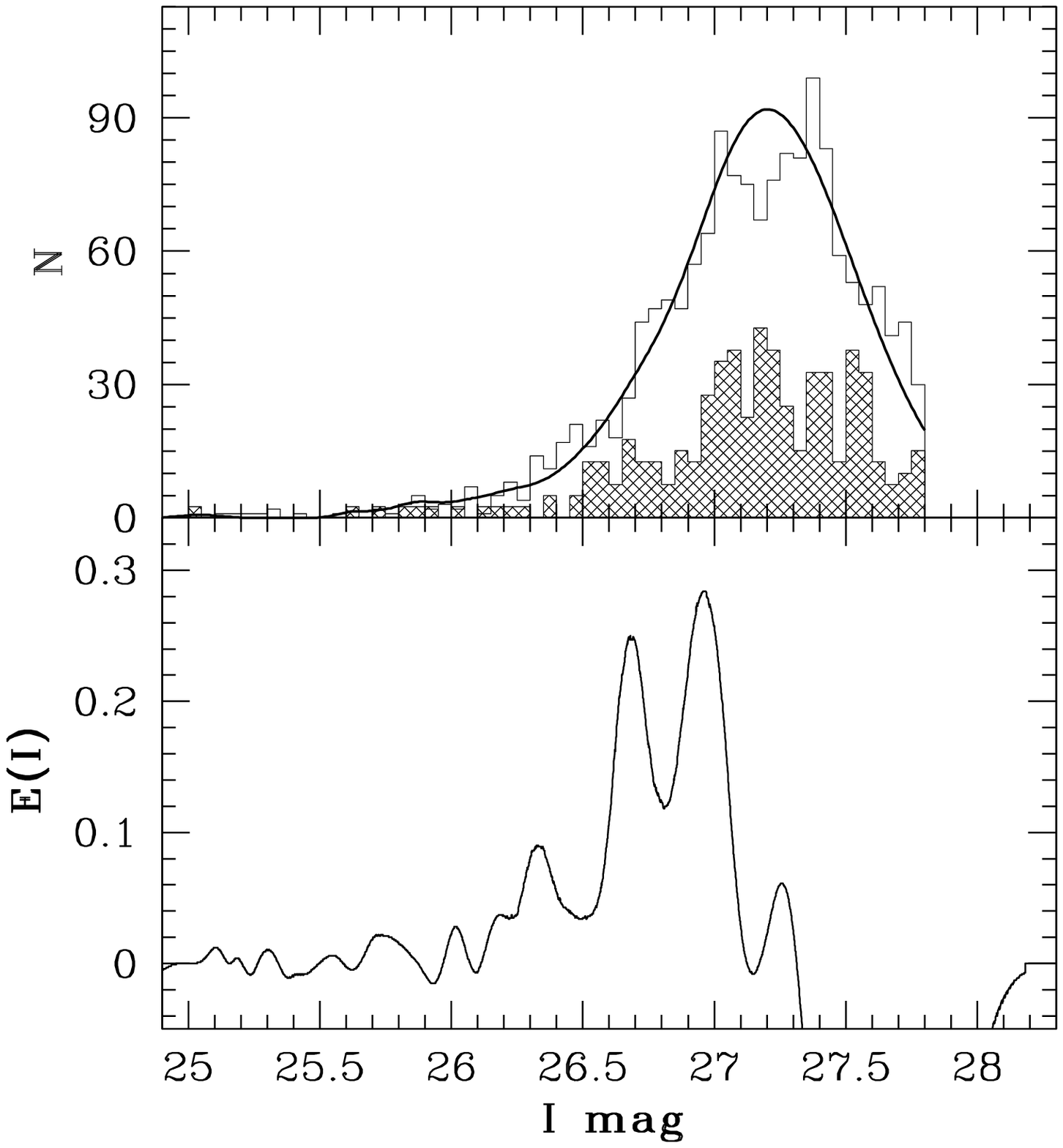}
\end{figure}

\clearpage
\begin{figure}
\epsscale{1.0}
\plotone{Harris.fig2.ps}
\end{figure}

\end{document}